\documentstyle[twocolumn,aps]{revtex}
\input epsf.sty

\def\SAS{\Delta_{\rm{SAS}}}
\def\CS{Chern-Simons }

\begin{document}

\title{ANOMALOUS STABILITY OF $\nu=1$ BILAYER QUANTUM HALL STATE}

\author{A. Sawada, Z.F. Ezawa, H. Ohno$^a$, Y. Horikoshi$^b$\\
O. Sugie, S. Kishimoto$^a$, F. Matsukura$^a$, Y. Ohno$^a$, M. Yasumoto} 

\address{
Department of Physics, Tohoku University, Sendai 980-77, Japan\\
${}^a$ Research Institute of Electrical Communication, 
Tohoku University, Sendai 980-77, Japan\\
${}^b$ School of Science and Engineering, Waseda University, 
Tokyo 169, Japan\\
}

\maketitle
\begin{abstract}
We have studied the fractional and integer quantum Hall (QH) effects 
in a high-mobility double-layer two-dimensional electron system.  
We have compared the "stability" of the QH state 
in balanced and unbalanced double quantum wells.
The behavior of the $\nu=1$ QH state is found to be strikingly 
different from all others.
It is anomalously stable, though all other states decay, 
as the electron density is made unbalanced between the two quantum wells. 
We interpret the peculiar features of the $\nu=1$ state
as the consequences of 
the interlayer quantum coherence developed spontaneously
on the basis of the composite-boson picture.\hfil\par

\end{abstract}

\section{Introduction}
The bilayer quantum Hall (QH) system possesses 
a rich phase diagram \cite{PhaseDiagr} since it allows 
three controllable parameters, the magnetic length $\ell _{\rm B}$, 
the interlayer distance $d$ and the symmetric-antisymmetric 
tunneling gap energy $\SAS$.  
By controlling these parameters we can realize 
various bilayer QH states $\Psi _{m_{\rm f}m_{\rm b}m}$ with $m_\alpha $ 
odd integers and $m$ an integer ($m_\alpha \geq m$)\cite{DBLYexp}.  
When the interlayer and intralayer Coulomb interactions becomes nearly equal, 
the $\Psi _{mmm}$ state is realized at $\nu \equiv 1/m$.  
A novel interlayer quantum coherence (IQC) has been predicted to develop 
spontaneously in the $\Psi_{mmm}$ states \cite{EIcoher},
which is characterized by a pseudo-gapless mode describing 
the interlayer phase and the electron density difference.
A typical QH state is given by $\Psi _{111}$ at $\nu =1$.  
The $\nu =1/m$ QH state is intriguing 
since there are two distinguishable states, 
the monolayer state stabilized by the tunneling interaction 
and the coherent state $\Psi_{mmm}$ stabilized by the Coulomb interaction.

We wish to address what are the signals for the spontaneous development
of IQC characterizing the $\Psi _{mmm}$ state.  
Murphy {\it et al.} \cite{Sheena} made an experiment 
in which they observed an anomalous activation energy dependence
of the $\nu=1$ state on the tilted magnetic field, 
which is probably one of the signals \cite{EIplasmon,YangMoon}.  
Furthermore, their data indicate that IQC is observable even 
at a large tunneling interaction, $\SAS \approx  8.4$K.

In search for a clear signal we have performed an experiment on 
bilayer QH states by varying the density ratio of the two layers
as well as the total electron density.
We hereafter refer to the QH states as "balanced QH states" 
when the densities are equal, and as "unbalanced QH states" 
when they are not equal.
Experiments have so far been made extensively 
on balanced QH states \cite{DBLYexp,DBLYexpB},
and only limited works have been done on the unbalanced QH states
\cite{Shayegan,Hamilton}.
The study of unbalanced QH states reveals its character:  
For instance, if the QH state is stabilized by the energy gap $\SAS$,
it decays as the system becomes unbalanced.

In our experimental data, all QH states depend sensitively 
on the density ratio except the $\nu=1$ state.
They have strong tendency to decay 
as the electron density between the two quantum wells is made unbalanced.

On the contrary, 
the behavior of the $\nu=1$ state is strikingly different from all others.
The $\nu=1$ state is stable even if the density ratio is changed, 
as far as the total density is less than a critical value.
We interpret that this stability of the $\nu=1$ state is an evidence 
of the IQC developed spontaneously.

\section{Experimental Results}
The sample was grown by molecular beam epitaxy on (100)-oriented 
GaAs substrate, and consists of two modulation-doped 200\,\AA\ 
quantum wells, separated by Al$_{0.3}$Ga$_{0.7}$As barriers of thickness 
$d_{\rm B}=31$\,\AA.  It has the total density 
$2.3\times 10^{11}\,$cm$^{-2}$ and the mobility 
$3.0\times 10^5\,$cm$^2$/Vs at 30\,mK at zero gate voltage.  
Measurements were performed with the sample mounted 
in the mixing chamber of a dilution refrigerator 
with a base temperature less than 6\,mK. 
The maximum field of our superconducting magnet is 
13.5\,T at 4.2\,K. Temperature was measured by a RuO$_2$ resistance 
set near the sample and calibrated by $^3$He melting pressure thermometer. 
Standard low-frequency ac lock-in techniques 
were used with currents less than  100 nA to avoid heating effects.

\begin{figure}[t]
\epsfxsize=75mm
\centerline{\epsfbox{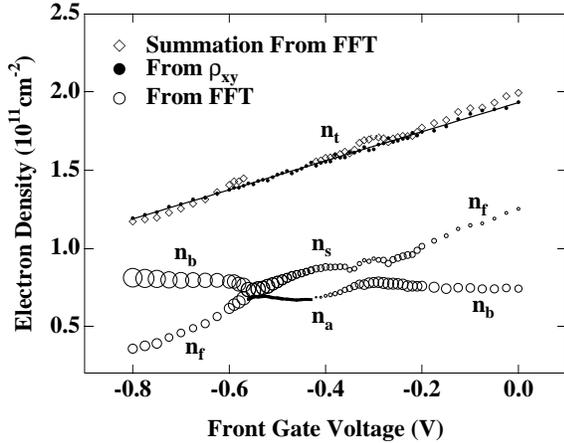}}
\caption{
Carrier density as a function of the front gate voltage. 
The open circles are the electron densities obtained from 
Fourier transforms of SdH signals.
The total electron densities (solid circles) are measured 
by the low field Hall resistance.
The size of the open circle is proportional to the signal intensity.
The open square is the sum of electron densities.
The solid curve in the density $n_{\rm a}$ is not observed by SdH
but is calculated by $n_{\rm a}=n_{\rm t}-n_{\rm s}$.  
}
\label{fig:1}
\end{figure}

Schottky front and back  gates were used to 
change the total carrie density and the density ratio 
of the two quantum wells.  
In Fig.\ref{fig:1} the densities $n_{\rm f}$ and $n_{\rm b}$ 
of the front and back layers were obtained 
from Fourier transforms of the low-field ($B<1.3$\,T) 
Shubnikov-de Haas (SdH) oscillations, 
and the total density $n_{\rm t}$ 
was obtained from the Hall resistance at low fields.  
The sum of $n_{\rm f}$ and $n_{\rm b}$ is in good agreement 
with $n_{\rm t}$ obtained from the Hall resistance.  
We note the following features:  
(A) As the front  gate voltage $V_{\rm fg}$ is increased, 
$n_{\rm f}$ rises for gate voltage $V_{\rm fg}>-0.2$\,V and 
$V_{\rm fg}\,<-0.6$\,V, and $n_{\rm b}$ slightly decreases 
due to a negative compressibility \cite{NegatPress}.  
(B) For the front gate voltage -0.6V$<V_{\rm fg}<-0.2$\,V the electrons 
are not localized in individual wells but are resonating between them
with the averaged densities $n_{\rm s}$ and $n_{\rm a}$ on 
the symmetric and antisymmetric states. 
The maximum density difference at $V_{\rm fg}=-0.41$\,V 
gives $\SAS\approx 6.8$\,K.  

\vskip10mm

\begin{figure}[t]
\epsfxsize=75mm
\centerline{\epsfbox{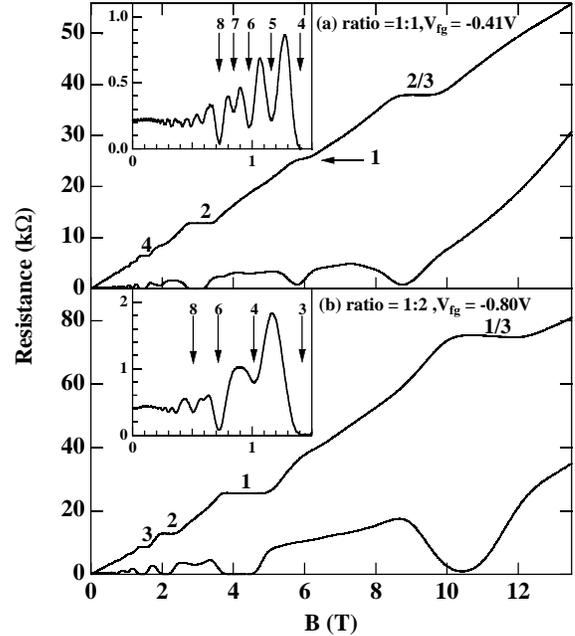}}
\caption{
Magneto and Hall resistance, $\rho_{xx}$ and $\rho_{xy}$, 
for two typical density ratios $n_{\rm f}/n_{\rm b}$.  
The appearance of QH states depends critically on this ratio.  
In (a) $n_{\rm f}=n_{\rm b}=6.9\times 10^{10}$cm$^{-2}$.
In (b) $n_{\rm f}=3.4\times 10^{10}$\,cm$^{-2}$ 
and $n_{\rm b}=7.4\times 10^{10}$cm$^{-2}$.
}
\label{fig:2}
\end{figure}

In Fig.\ref{fig:2} we show the magneto- and Hall resistance 
at $V_{\rm bg}=-37.2$V under two typical $V_{\rm fg}$.
Balanced QH states are seen in Fig.\,2(a) at $V_{\rm fg}=-0.41$V,
where $n_{\rm f}/n_{\rm b}=1/1$.
Unbalanced QH states are seen in Fig.\,2(b) at $V_{\rm fg}=-0.80$V, 
where $n_{\rm f}/n_{\rm b}\approx1/2$.

Odd-integer QH states in the inset of Fig.\ref{fig:2}(a) 
are the monolayer states stabilized by the tunneling interaction.
It exists only up to a certain magnetic field, 
beyond which bilayer QH states stabilized by the Coulomb interaction
are expected to occur \cite{PhaseDiagr}.  
The $\nu =1$ QH state, which is unstable at higher total electron density, 
has been stabilized at $n_{\rm t}=1.4\times10^{11}$ cm$^{-2}$. 
This property can be understood based on the phase diagram \cite{PhaseDiagr}. 
On the other hand,
even-integer QH states will be bilayer states made of the same monolayer
QH states localized in the two wells,
which we call compound states with 
$(\nu_{\rm f},\nu_{\rm b})=(\nu/2,\nu/2)$,
or the monolayer state stabilized by the tunneling interaction.  
We cannot distinguish them.
They are present since these states are robust.
The $\nu =2/3$ state is the $\Psi_{330}$ state which is a compound state 
with $(\nu_{\rm f},\nu_{\rm b})=(1/3,1/3)$.

In Fig.\ref{fig:2}(b) 
we report the QH effect in the unbalanced configurations 
obtained by controlling the bias voltages.
The $\nu =3$ and $6$ states are compound states 
with $(\nu _{\rm f},\nu _{\rm b})=(1,2)$ and $(2,4)$, respectively,  
where $\nu=\nu_{\rm f}+\nu_{\rm b}$  and
$\nu_{\rm f}/\nu_{\rm_b}=n_{\rm f}/n_{\rm b}$.
The $\nu =4$ and 8 states are the "tails" of the compound states 
with  $(\nu _{\rm f},\nu _{\rm b})=(1,3)$ and $(2,6)$
broaden by disorder in the sample,
because the plateau became wider as the system approached 
$n_{\rm b}/n_{\rm f}=1/3$.
The origin of the $\nu=1$ and $1/3$ state is discussed 
in the following stage.

\begin{figure}[t]
\epsfxsize=75mm
\centerline{\epsfbox{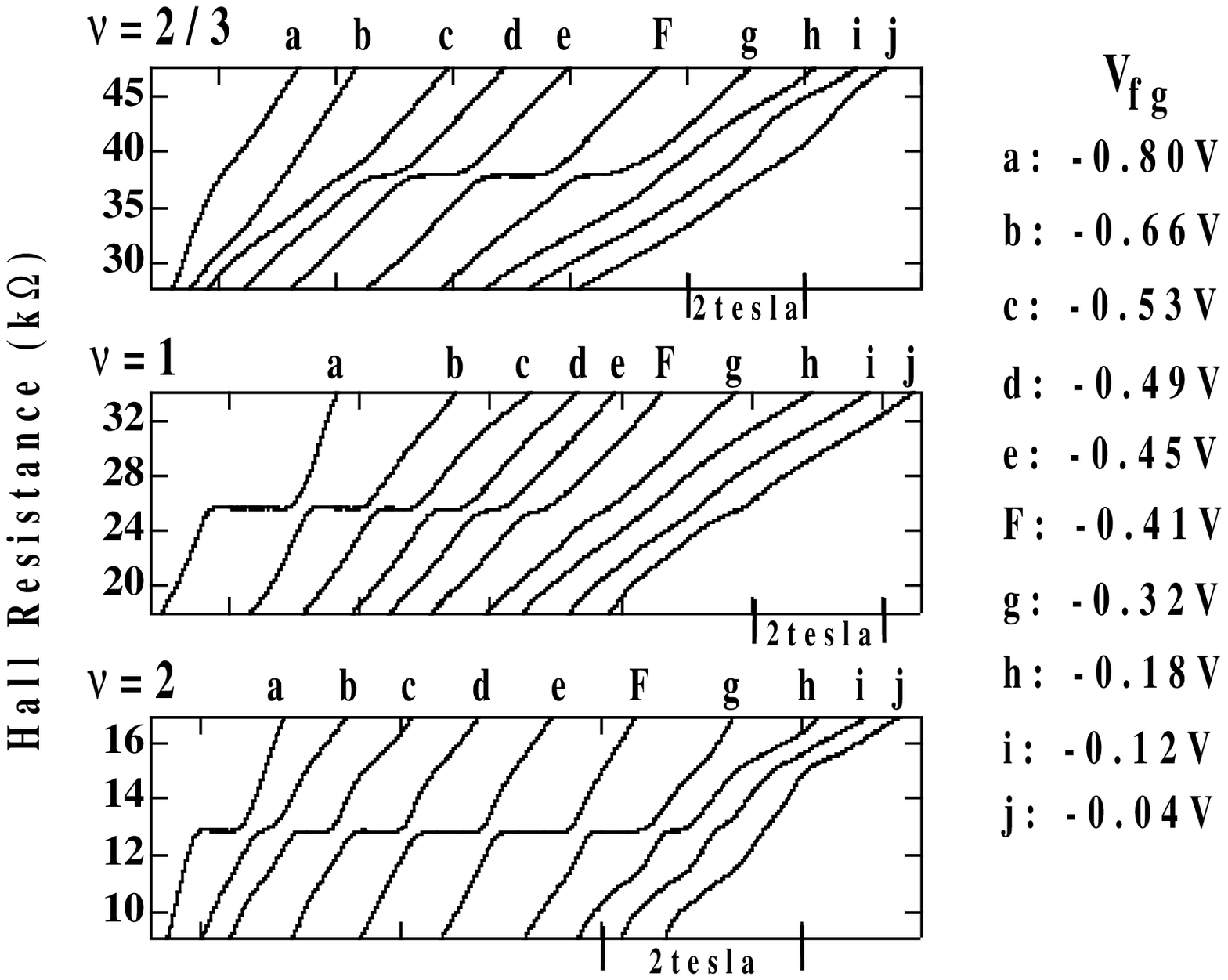}} 
\caption{
The Hall resistances near the plateau of 
$\nu $=2/3, 1 and 2 QH state are shown at various $V_{\rm fg}$.  
The curve (F) represents balanced QH effect.
}
\label{fig:3a}
\end{figure}

\begin{figure}[t]
\epsfxsize=70mm
\centerline{\epsfbox{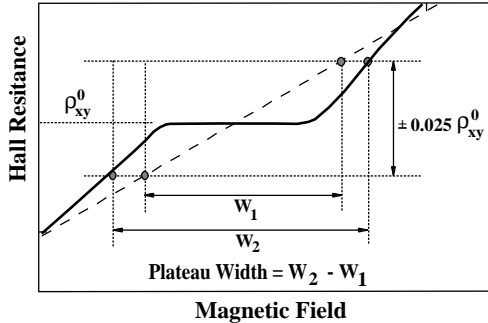}} 
\caption{
The definition of the Hall plateau width.
The dotted line represents a classical Hall resitance,
and $\rho_{xy}^{0}$ is a quantized Hall resistance.
}
\label{fig:5}
\end{figure}

In Fig.\ref{fig:3a} we give the Hall resistance measured at 
fixed $V_{\rm bg}=-37.2$\,V and various $V_{\rm fg}$.
It is customary to discuss the stability of 
the QH states by the value of the magnetoresistance ($\rho_{xx}$).  
We propose to discuss the stability in terms of the width of the Hall plateau. 
We have defined the width of the Hall plateau by the width of 
the magnetic field within the $\pm2.5$\,\% range of 
the Hall resistance after subtracting the classical Hall resistance,
as illustrated in Fig.\ref{fig:5}.

\begin{figure}[t]
\epsfxsize=80mm
\centerline{\epsfbox{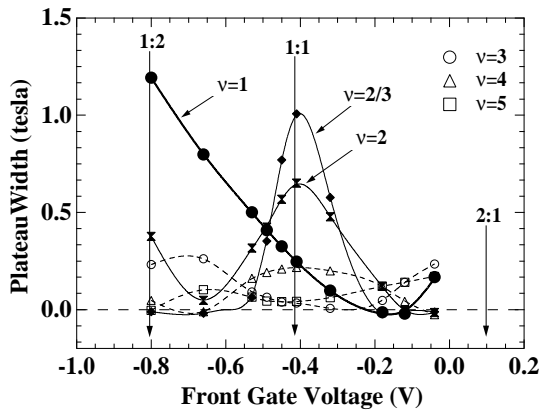}}
\caption{
The Hall plateau width as a function of the front gate voltage 
$V_{\rm fg}$.  The curved lines interpolate data points. 
}
\label{fig:3b}
\end{figure}

As is seen in Fig.\ref{fig:3b}, 
the plateau widths of the $\nu =2$ and $2/3$ state are 
widest when $n_{\rm f}/n_{\rm b}=1$, 
and decrease monotonically as it deviates from the balanced point.  
This is explained by considering that
the $\nu =2$ and $2/3$ states are compound states 
with $(\nu _{\rm f},\nu _{\rm b})=(1,1)$ and (1/3,1/3), respectively. 
The $\nu =2$ data show an increase below $-0.7$V, 
and reaches a maximum at $-0.8$V 
which is the edge of our experimental region.  
We interpret it as a tail of the monolayer state 
with $(\nu _{\rm f},\nu _{\rm b})=(0,2)$.
In the figure the behavior of the $\nu=1$ QH state is strikingly
different from all the other states.
The state is stable over a wide range of density ratio,
and at the same time no peak is observed at the balanced point.
The QH state would be stablest at the balance point if it were
stabilized by either $\SAS$ or any simple correlations.

\begin{figure}[t]
\epsfxsize=75mm
\centerline{\epsfbox{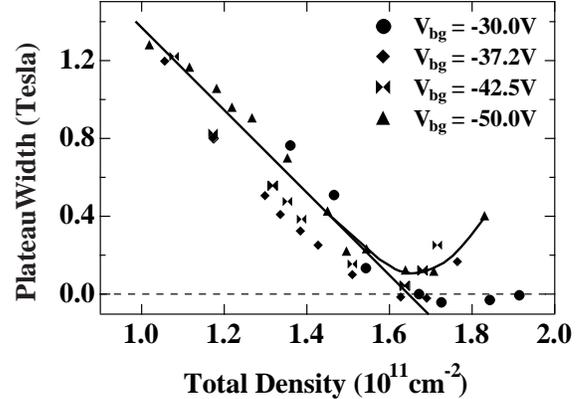}}
\caption{
Stability of $\nu=1$ as a function of the total density.
The dependence of the Hall plateau width on the total density 
is given for four different back gate voltage $V_{\rm bg}$.
Almost all data points are on a straight line 
blow the density $1.6\times10^{11}$\,cm$^{-2}$, 
and deviate from it by the tail of 
the $(\nu _{\rm f},\nu_{\rm b})=(2/3,1/3)$ state.
}
\label{fig:4}
\end{figure}

In Fig.\ref{fig:4} the width of the Hall plateau of the $\nu=1$ state 
is given as a function of 
$n_{\rm t}$ for four different back gate voltage $V_{\rm bg}$. 
It is seen that the plateau width depends 
on the total density $n_{\rm t}$ but not on the density ratio,  
suggesting an interesting scaling law.
The stability decreases as $n_{\rm t}$ increases, and
the QH state becomes unstable 
when $n_{\rm t} \approx 1.7\times10^{11}$\,cm$^{-2}$.
As the density increases beyond $1.7\times10^{11}$\,cm$^{-2}$,
the $\nu=1$ state becomes stable again.
This increase is probably due to the tail of the compound state with
$(\nu _{\rm f},\nu_{\rm b})=(2/3,1/3)$ having the peak 
at $n_{\bf f}/n_{\bf b}=2/1$. 

\section{Interlayer Quantum Coherence}
In this section we discuss the origin of the peculiar property of the $\nu=1$
QH state.  
We relate it to IQC based on the composite-boson picture 
\cite{EIcoher,EIplasmon,EzaIQC}.

In the bilayer system the Hamiltonian $H$ is the sum of 
the kinetic term $H_{\rm K}$, 
the Coulomb term $H_{\rm C}$ and the tunneling term $H_{\rm T}$.
(For simplicity we freeze the spin degree of freedom.)  
The Coulomb term may be written as a sum of two terms \cite{EzaIQC},
$H_{\rm C}=H_{\rm C}^{+} + H_{\rm C}^{-}$, 
where $H_{\rm C}^{+}$ depends only on the 
total electron density $n_{\rm t}$
while $H_{\rm C}^{-}$ depends only on the 
density difference $n_{\rm f}-n_{\rm b}$.
The term $H_{\rm C}^{-}$ describes  
the capacitance energy stored between the two layers.
The key behavior is that
$H_{\rm C}^{-}\rightarrow 0$ as $d/\ell _{\rm B}\rightarrow 0$ and 
$H_{\rm C}^{-}\rightarrow H_{\rm C}^{+}$ as 
$d/\ell _{\rm B}\rightarrow \infty $.  
The system has a rich phase diagram depending on relative strength of these 
interaction terms.
We expect to have the following three phases (A) $\sim$ (C).

When the capacitance term $H_{\rm C}^{-}$ is negligible,
the bilayer system is mapped exactly to a monolayer system with the spin 
degree of freedom, where the tunneling term is mapped to the Zeeman term  
\cite{YangMoon,EzaIQC}. 
Therefore, we have two phases:
(A) For a large $\SAS$, the bilayer QH state is reduced to a "monolayer" QH 
state built on the symmetric or antisymmetric state.
It is clear that the density ratio is fixed as $n_{\rm f}/n_{\rm b}=1$
for the state to realize.
(B) For a small $\SAS$ it is considered as a QH ferromagnet 
\cite{YangMoon,EzaIQC} identified with the $\Psi_{mmm}$ state, 
about which we explain soon.

The capacitance term $H_{\rm C}^{-}$ acts to break these phases,
yielding a new phase (C) described by the $\Psi_{m_{\rm f}m_{\rm b}m}$ state
with $m_{\rm f}m_{\rm b}\not=m^2$.  
This phase realizes \cite{EIcoher} at
\begin{equation}
\nu  ={m_{\rm f}+m_{\rm b}-2m \over  m_{\rm f} m_{\rm b}-m^2} ,
\end{equation}
where the density ratio is fixed as
\begin{equation}
{n_{\rm f}\over n_{\rm b}}=
{m_{\rm b}-m\over m_{\rm f}-m} .
\label{FixedDensi}
\end{equation}
A special limit with $m=0$ corresponds to the compound state
with $(\nu_{\rm f},\nu_{\rm b})=(1/m_{\rm f},1/m_{\rm b})$.
It should be emphasized that the QH state exists only 
at a fixed density ratio in phases (A) and (C).

We now explain why phase (B) is peculiar.
The nature of the QH ferromagnet 
is most easily revealed based on the 
composite-boson picture.
The composite-boson field is defined with the aid of the \CS (CS) field.  
We first neglect the terms $H_{\rm C}^-$ and $H_{\rm T}$ in the Hamiltonian. 
Since the term $H_{\rm C}^{+}$ does not 
discriminate electrons belonging to different layers, 
only one CS field $\bbox{C}$ is introduced.
The CS flux is attached to each electron and determined by
\begin{equation}
\hbox{rot\ } \bbox{C}=m n_{\rm t} \phi_0  ,
\label{CSconstSPN} 
\end{equation}
in terms of the total density $n_{\rm t}$
and the Dirac flux unit $\phi_0=2\pi\hbar c/e$.  
Bose condensation occurs when the 
external magnetic field is cancelled by the CS field, 
$\bbox{C}+\bbox{A}^{\rm{ext}}=0$.  The 
constraint (\ref{CSconstSPN}) dictates that 
this is only possible at the filling 
factor
\begin{equation}
\nu ={1\over m}.  
\end{equation}
The essential point is that 
{\it the density ratio $n_{\rm f}/n_{\rm b}$ is arbitrary}.  
Namely, we obtain degenerate ground states $|\Delta{n}\rangle$ with
arbitrary density difference $\Delta{n}=n_{\rm f}-n_{\rm b}$.  
The interlayer phase $\theta$ is the variable conjugate to $\Delta{n}$.
The coherent states read
\begin{equation}
|\Psi(\theta)\rangle= \sum_{\Delta{n}}e^{-i\theta\Delta{n}} |\Delta{n}\rangle.
\label{CoherState}
\end{equation}
The coherent mode is a Goldstone mode.
Although the degeneracy is removed by 
the capacitative term $H_{\rm C}^{-}$ and the tunneling term $H_{\rm T}$,
there are states having various density difference $\Delta{n}$ 
and hence the coherent states (\ref{CoherState}) persist. 
The coherent mode is made gapful and called a pseudo-Goldstone mode.
Various IQC phenomena are expected to occur on these states
\cite{EIcoher,EIplasmon,YangMoon,EzaIQC}.

As we have explained,
the condition for the emergence of IQC is
the existence of QH states with arbitrary density difference $\Delta{n}$. 
In our data only the $\nu=1$ QH state has a peculiar feature that 
it continues to exist over a wide range of $\Delta{n}$.
Another characteristic feature of the $\nu=1$ QH state is that 
the stability depends only on the total density 
and not on the density difference.  
This is naturally explained in the composite-boson picture. 
Since the $\Psi_{111}$ state is considered as an integer QH state 
made of electrons disregarding their layer dependence 
as in (\ref{CSconstSPN}),
the stability should not depend on the density difference.
Furthermore,
as the density becomes larger, the magnetic length $\ell _B$ 
becomes smaller.
Since the ratio $d/\ell_B$ becomes larger,
the interlayer Coulomb interaction becomes effectively smaller. 
It acts to break the $\Psi_{111}$ state.  
Hence, the stability decreases as the total density increases,
and the QH state breaks down eventually at a critical density.

In conclusion the best signal for the emergence of IQC 
is the existence of QH states $|\Delta{n}\rangle$ with 
various density difference $\Delta{n}$.
We have observed these QH states in unbalanced double quantum wells 
by applying bias voltages to the system.

\section{Acknowledgement}
We thank T. Saku (NTT) for providing us the sample used 
in the present work, to K. Hirakawa and S. Kawakami for providing us 
samples used in the early stage,
to O. Sakai and T. Nakajima for useful discussions,
and M. Suzuki for technical assistance.
This work was partly done at Laboratory for Electronic Intelligent Systems, 
Research Institute of Electrical Communication, Tohoku University.
This work was supported by grants from the Ministry of Education,
Science, Sports and Culture of Japan, 
and from Multi-disciplinary Science Foundation.



\end{document}